\documentclass{article}

\usepackage[preprint,nonatbib]{neurips_2023}

\usepackage[utf8]{inputenc} 
\usepackage[T1]{fontenc}    
\usepackage{hyperref}       
\usepackage{url}            
\usepackage{booktabs}       
\usepackage{amsfonts}       
\usepackage{nicefrac}       
\usepackage{microtype}      
\usepackage{xcolor}         
\usepackage{graphicx}
\usepackage{bm}
\usepackage{amsmath}
\usepackage{amssymb}
\usepackage{subfigure}
\usepackage{caption}
\usepackage{multirow}
\usepackage{makecell}
\usepackage{wrapfig}
\usepackage[resetlabels]{multibib}
\newcites{app}{Appendix Reference}

\title{Large-Scale Cell Representation Learning via Divide-and-Conquer Contrastive Learning}

\author{%
    \textbf{Suyuan Zhao}$^{1, 2}$\thanks{\quad Equal contribution.}\enspace\thanks{\quad Work done during Suyuan Zhao’s internship at AIR, Tsinghua University.}\enspace, 
    \textbf{Jiahuan Zhang}$^{1}$\footnotemark[1]\enspace, 
    \textbf{Zaiqing Nie}$^{1}$\thanks{\quad Corresponding author.}\\
    Institute for AI Industry Research (AIR), Tsinghua University$^{1}$\\
    Department of Computer Science and Technology, Tsinghua University$^{2}$ \\
    \texttt{sxdtzsy@gmail.com};\enspace\texttt{\{zhangjiahuan, zaiqing\}@air.tsinghua.edu.cn} \\
}

\hypersetup{
colorlinks=true,
linkcolor=black
}

\begin{document}

\maketitle

\begin{abstract}
Single-cell RNA sequencing (scRNA-seq) data is a potent tool for comprehending the ``language of life'' and can provide insights into various downstream biomedical tasks. Large-scale language models (LLMs) are starting to be used for cell representation learning. However, current LLM-based cell representation learning methods depend solely on the BERT architecture, causing an anisotropic embedding space that leads to inefficient semantic representation. Contrastive learning alleviates this problem by distributing the embeddings uniformly. As a larger batch size in contrastive learning results in better representation, the practical application of contrastive learning in cell representation learning is hampered by the high dimensionality of scRNA-seq data and the large parameter volume of LLMs. To address the batch size limitation, we propose a novel divide-and-conquer contrastive learning approach to decouple the batch size from the GPU memory size for cell representation learning. Based on our divide-and-conquer contrastive learning approach, we introduce Single-\textbf{Cell} \textbf{L}anguage \textbf{M}odel (\textbf{CellLM}), a large-scale cell representation learning model to handle high-dimensional scRNA-seq data with tens of thousands of genes. CellLM has over 50 million parameters trained with 2 million scRNA-seq data and makes the first attempt to learn cell language models from both normal cells and cancer cells. CellLM achieves new state-of-the-art (SOTA) results in all evaluated downstream tasks: including a 71.8 $F_1$-score for cell type annotation (a 3.0\% absolute improvement over scBERT), an average $F_1$-score of 88.9 for single-cell drug sensitivity prediction in a few-shot scenario (an 8.3\% absolute improvement), and a 93.4 Pearson's correlation for single-omics cell line drug sensitivity prediction (a 6.2\% absolute improvement). The pre-trained model, codes, and datasets for our CellLM are accessible at \url{https://github.com/BioFM/OpenBioMed}.

\end{abstract}

\section{Introduction}

Single-cell RNA sequencing (scRNA-seq) data is a potent tool for comprehending the ``language of life'' and can provide insights into various downstream biomedical tasks. Large-scale language models (LLMs) are beginning to be used to decipher the coding language of life and have achieved some success \cite{mo2021multimodal, DNABERT_ji_2021}, and several very recent studies have shown the effectiveness and feasibility of applying LLMs for the representation of single-cell data \cite{ ScBERT_yang_2022, Connell_2022_neruips}. scBERT \cite{ScBERT_yang_2022} is the first study to encode scRNA-seq data using the LLM approach. It uses more than one million normalized unlabeled scRNA-seq data, and utilizes the BERT-based pre-training model, Performer, to obtain the representation of scRNA-seq data. Exciver \cite{Connell_2022_neruips} works in a similar manner. 

However, these approaches depend solely on the BERT architecture for cell representation, and studies \cite{BERT_flow_li-etal-2020, Sentence-bert_reimers_gurevych_2019} have revealed that directly applying BERT may lead to a degradation in its representation quality due to the anisotropy of the embedding space. More specifically, low-frequency words are not effectively trained during pre-training, causing their embedding vectors to be sparsely distributed in the feature space. On the other hand, the embedding vectors of high-frequency words, which are well-trained, tend to cluster together in the feature space. The uneven distribution of the embedding space restricts the ability to measure semantic associations between high-frequency and low-frequency words. Likewise, scRNA-seq data show diverse gene expression frequencies, indicating that the shortcomings of the BERT architecture will carry over to the semantic representation process of scRNA-seq data.

Contrastive learning addresses the anisotropy issue by uniformly distributing embeddings through learning positive and negative sample features. However, a fundamental challenge in contrastive learning lies in ensuring a sufficient number of negative samples in each training batch, which is vital for the model to learn effective features. Unfortunately, this challenge is exacerbated by the GPU memory size limitation, making it difficult to significantly increase the batch size, especially in the most common end-to-end contrastive learning methods \cite{MOCO_He2019}. 

In the case of scRNA-seq data, the data dimension is determined by the number of genes in a cell, which can be as high as 19,379 known protein-coding genes in humans, even higher if non-coding genes are included \cite{HGNC_Seal2022GenenamesorgTH}. The high dimensionality of scRNA-seq data and the enormous parameters of LLMs pose a challenge in achieving a large batch size when using contrastive learning for cell representation tasks, especially when the size of GPU memory is restricted. Therefore, it is crucial to decouple the batch size from the GPU memory size.

The existing methods have made strides in tackling these challenges. For instance, the memory bank approach \cite{Memory_bank} addresses this issue by expanding the number of negative samples through the maintenance of a vast negative sample queue that is not involved in gradient backpropagation. However, this method results in asynchronous updates of encoders for positive and negative samples, causing discrepancies not only between the samples but also between the encoders, resulting in training instability. MoCo \cite{MOCO_He2019} proposes a momentum encoder to mitigate inconsistent updates between the negative encoder and the positive encoder in the memory bank. However, in comparison to the end-to-end contrastive learning approach, MoCo only reduces the asynchronous update of the encoder. The truth of the matter is that the embeddings of positive and negative samples are still produced by different encoders.

To overcome the limitation of batch size and ensure that positive and negative samples are generated by the same encoder, we propose a novel \textbf{divide-and-conquer contrastive learning} approach to decouple the batch size from the GPU memory size for cell representation learning. More importantly, the divide-and-conquer contrastive learning method has been mathematically rigorous and has been proven to be completely equivalent to the end-to-end contrastive learning method. It means that we can increase the batch size while simultaneously updating the encoder for positive and negative samples without introducing any additional errors. By leveraging the concept of trading time for space, a big batch is divided into several smaller mini-batches. The gradient update calculations are then carried out in sequence, allowing us to increase the appropriate batch size without compromising the synchronization of encoder updates for positive and negative samples. By distributing the workload over multiple smaller batches, we can effectively utilize available GPU memory resources while maintaining consistency in the encoder updates.

Based on our divide-and-conquer contrastive learning approach, we introduce Single-\textbf{Cell} \textbf{L}anguage \textbf{M}odel (\textbf{CellLM}), a large-scale cell representation learning model to handle high-dimensional scRNA-seq data with 19,379 genes. CellLM has over 50 million parameters trained with 2 million scRNA-seq data. It’s worth mentioning that CellLM is the first attempt to learn cell language models from both normal and cancer cells. Since cancer scRNA-seq data is helpful in understanding and treating cancer at the single-cell level \cite{Li_cancer}, it ultimately leads to more widespread and cost-effective treatment options at the human body level. In addition, due to the sparsity of single-cell data, we reduce the computational load of pre-training by dynamically incorporating genes with expressions, instead of utilizing full-length gene sequences.

We validate the performance of CellLM on a range of downstream biomedical tasks and achieve new SOTA in all evaluated tasks: including a 71.8 $F_1$-score for cell type annotation (a 3\% absolute improvement over scBERT), an average $F_1$-score of 88.9 for single-cell drug sensitivity prediction in a few-shot scenario (an 8.3\% absolute improvement), and a 93.4 Pearson's correlation for single-omics cell line drug sensitivity prediction (a 6.2\% absolute improvement). Our experiments demonstrate that CellLM produces a superior representation of single-cell, enhances the semantic representation of cell line data, and enables more precise virtual drug screening along the entire chain.

Our main contributions can be summarized as follows:
\begin{itemize}

\item We propose a novel divide-and-conquer contrastive learning approach to decouple the batch size from the GPU memory size for cell representation learning, which effectively addresses the embedding space anisotropy problem caused by the BERT architecture. The divide-and-conquer contrastive learning has undergone strict mathematical analysis and has been proven to be completely equivalent to the end-to-end contrastive learning method.

\item We introduce CellLM, a large-scale cell representation learning model with over 50 million parameters trained with 2 million scRNA-seq data. CellLM makes the first attempt to learn cell language models from both normal cells and cancer cells. 

\item CellLM achieves SOTA results on a range of downstream biomedical tasks. It has been proven to enhance cell representation and aid in virtual drug screening.
\end{itemize}

\section{Related works}

\paragraph{Representation of scRNA-seq data.}

The gene expression profile provides valuable information about the expression levels of genes within a single cell, making it a crucial component of research studies. Nearly 20k human protein-coding genes are known now \cite{HGNC_Seal2022GenenamesorgTH}, and directly analyzing such high-dimensional data poses an extreme challenge. Additionally, scRNA-seq data suffer from a high false dropout rate, leading to the ``Dropout Zeros'' phenomenon \cite{DZ2_Svensson2019DropletSI, DZ3_Silverman2018NaughtAZ, DZ1_Linderman2022ZeropreservingIO}. Researchers have devised a range of methods to tackle the challenges of such highly sparse and noisy data.

Traditional approaches attempt to analyze scRNA-seq by dimensionality reduction, such as manually selecting the marker genes \cite{Pasquini2021AutomatedMF, Guo2021scSorterAC}, machine learning methods \cite{PCA_FRS1901LIIIOL, PCA_Shen2009PrincipalCA, PCA_Hotelling1933AnalysisOA, PCA_Tsuyuzaki2019BenchmarkingPC, tsne_Maaten2014AcceleratingTU}, or autoencoder-based methods \cite{AE_Alessandri2020SparselyconnectedA, AE_Talwar2018AutoImputeAB, AE_Tran2019FastAP, AE_Tran2022scCANSC}. However, the genes selected manually are often based on empirical observations  \cite{problem_Huang2020EvaluationOC}, and machine learning methods have high complexity and limited noise resistance. The autoencoder-based methods heavily rely on the similarity between test and training data.

LLMs have the potential to be applied to modeling scRNA-seq data. scBERT \cite{ScBERT_yang_2022} is the first work to encode scRNA-seq data by LLM. scBERT employs the Performer \cite{Performers_choromanski2022rethinking} architecture to obtain gene expression representations of single-cells from over a million normalized unlabeled scRNA-seq data with 6 million parameters. Compared with domain-specific tools, scBERT achieves superior performance on the cell type annotation task. Exceiver \cite{Connell_2022_neruips} uses the Perceiver IO \cite{Perceiver_IO_jaegle2022perceiver} architecture to obtain representations of single-cell count matrix data. It is pre-trained on almost 0.5 million count matrix data of single-cell gene expression from healthy humans, and its effectiveness is verified on downstream tasks.

\paragraph{Contrastive Learning.}

Contrastive learning (CL) is a widely utilized self-supervised learning technique in computer vision (CV) \cite{simclr, MOCO_He2019} and natural language processing (NLP) domains \cite{gao-simcse, consert}. It aims to train the encoder by generating similar embeddings for data in the same class while maximizing the dissimilarity between embeddings of different classes in the feature space. However, a significant challenge in contrastive learning is ensuring an adequate number of negative samples in each training batch, as it is crucial for the model to learn effective features. Unfortunately, this challenge is further complicated by the limitation of GPU memory size, particularly in popular end-to-end contrastive learning methods \cite{MOCO_He2019}. Numerous research efforts have focused on addressing this issue and proposed various solutions \cite{Memory_bank, MOCO_He2019}.

Several works have applied CL to improve the representation of scRNA-seq data. Concerto \cite{Concerto_Yang2022ContrastiveLE} employs a teacher-student network to construct positive samples in CL. CLEAR \cite{CLEAR_Han2021SelfsupervisedCL} uses data augmentation methods such as adding noise, random mask, inner swap, and dropout to construct positive samples. However, artificially perturbing the original data may alter the semantics of genes, which may not be a suitable data augmentation method for scRNA-seq data analysis where each gene expression level corresponds to a specific meaning.

\section{Methods}

\begin{figure}[htp]
  \centering
  \makebox[\textwidth][c]{\includegraphics[width=1\linewidth]{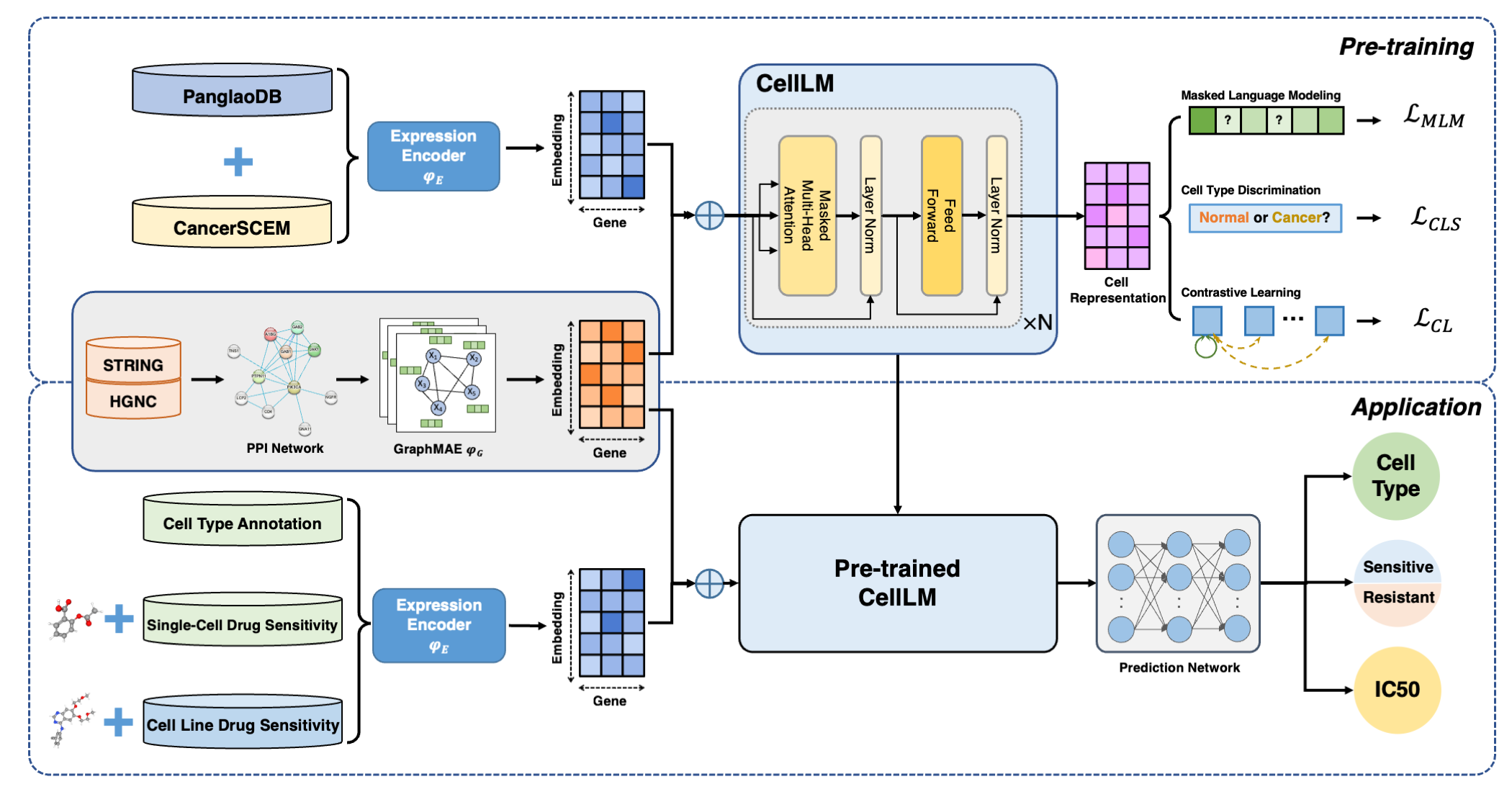}}
  \captionsetup{font={small}}
  \caption{The framework of the CellLM. \textbf{Pre-training:} The CellLM is pre-trained on PanglaoDB and CancerSCEM. The gene expression data and gene interaction data are encoded and then added together before being fed into the model for self-supervised learning. During the pre-training, we use three self-supervised tasks, masked language modeling (MLM), cell type discrimination, and a novel divide-and-conquer contrastive learning. \textbf{Application:} We evaluate the pre-trained CellLM on cell type annotation and drug sensitivity tasks at single-cell and cell line data levels.}
  \label{figure1}
\end{figure}

In this section, we provide a comprehensive description of the CellLM workflow. The framework of the CellLM is illustrated in Fig.\ref{figure1}.

\subsection{Pre-Processing of scRNA-seq Data}
The raw data of scRNA-seq is provided as a count matrix, denoted as $M=\{n_k\}_{k=1}^N$, where $n_k$ is an integer representing the count of the $k$-th gene in the scRNA-seq and N is the number of genes. Due to the variations in sequencing protocols and conditions, the data from different sequencing batches have different levels and are not comparable. So we normalize them to $X=\{x_k\}_{k=1}^N$, where
\[x_k=\log(1+\frac{n_k}{\sum_{j=1}^{N}n_j}\times 10000).\]
As scRNA-seq data is high dimensionality and sparsity, we only select the positions of each non-zero gene expression in the cell $P = \{p_k\}=\{j | x_j\in X\; and\; x_j\neq 0\}$, and their corresponding expression levels in $Y=\{y_k\}=\{x_{p_k}\}$. This reduces the original training cost by over 70\%. 

\subsection{Model Architecture}
We introduce the architecture of CellLM, it consists of three trainable parts: 

\paragraph{Expression encoder ($\varphi_E$).}
The gene expression level is obtained by normalizing the count matrix and is actually discrete in each cell. Ignoring this characteristic and treating it as continuous for model input will introduce significant noise. Therefore, we divide it into several bins based on expression level and map each bin to a trainable 512-dimensional encoding.

\paragraph{Gene Encoder ($\varphi_G$).}
The protein-protein interaction (PPI) network can reflect the relationships between genes. As shown in Fig.\ref{figure1}, we use one of the state-of-the-art graph representation learning methods, GraphMAE \cite{graphMAE}, to obtain gene embeddings as additional knowledge.

\paragraph{Performer-based module.}
The Performer model is a variant of the Transformer model. In the Transformer model, the complexity of computing the attention matrix is quadratic in the sequence length, while the Performer calculates an approximate attention matrix with linear complexity. Since the length of the scRNA-seq is relatively large, using the Performer architecture can significantly reduce memory usage and improve computational efficiency.

We input $\{P, Y\}$ obtained from Section 3.1 into the model. The input matrix $C=\{c_k\}$ to the Performer consists of two parts: gene embedding $\varphi_G(p_k)$ and expression embedding $\varphi_E(y_k)$, we inject knowledge of gene interactions for each gene expression by adding them. 
$C$ is then passed through the Performer model to obtain the encoding $H=\{h_k\}$. That is,
\begin{gather*}
    c_k = \varphi_G(p_k)+\varphi_E(y_k) \\
    H = \mathrm{Performer}(C) \\
\end{gather*}

\subsection{Pre-training CellLM}
During the pre-training process, researchers use suitable self-supervised tasks to help the model learn general representations of the data, which can be widely applied or transferred to multiple downstream applications. In order to improve the cell representation capability of the model, we set the following three self-supervised learning methods based on the features of the scRNA-seq data we used.

\subsubsection{Divide-and-Conquer Contrastive Learning}

We introduce contrastive learning to alleviate the problem of representation degradation caused by BERT-based methods and enhance the encoding capability for cell representation. A crucial step in contrastive learning is constructing positive and negative samples for data augmentation, which helps the model better understand the data features. In scRNA-seq data, each gene expression level carries unique meaning, and artificial data augmentation methods such as shuffling and perturbing at the input data may disrupt the gene expression semantics. We believe that perturbation at the feature level is more suitable for data augmentation in scRNA-seq data. Therefore, we use two instances of standard dropout applied to the same single-cell to construct positive samples \cite{gao-simcse}, while other single-cells in the same batch serve as negative samples. The loss function is as follows:
\[ \mathcal{L}_{CL}=-\frac{1}{T}\sum_{i=1}^{T}\log{\frac{\mathrm{e}^{\mathrm{sim}(\bm{h}_i, \bm{h}_i^+)/\tau}}{\sum_{j=1}^{T}\mathrm{e}^{\mathrm{sim}(\bm{h}_i, \bm{h}_j^+)/\tau}}}, \]
where $T$ is the batch size, $\mathrm{sim}$ is the cosine similarity function, $\tau$ is the temperature parameter, $\bm{h}_i$ and $\bm{h}_i^+$ represent the encoding results for the $i$-th data and its positive sample, respectively.

Due to the enormous parameters of the model and high-dimensionality input sequences, it is challenging to train the model with a large batch size within a limited GPU memory size. However, contrastive learning requires a large batch size to provide enough negative samples. Therefore, we designed divide-and-conquer contrastive learning to separate the batch size from the actual amount of data being simultaneously processed. 

\begin{wrapfigure}{r}{0.55\linewidth}
    \vspace{-0.85cm}
     \begin{center}
      \includegraphics[width=0.85\linewidth]{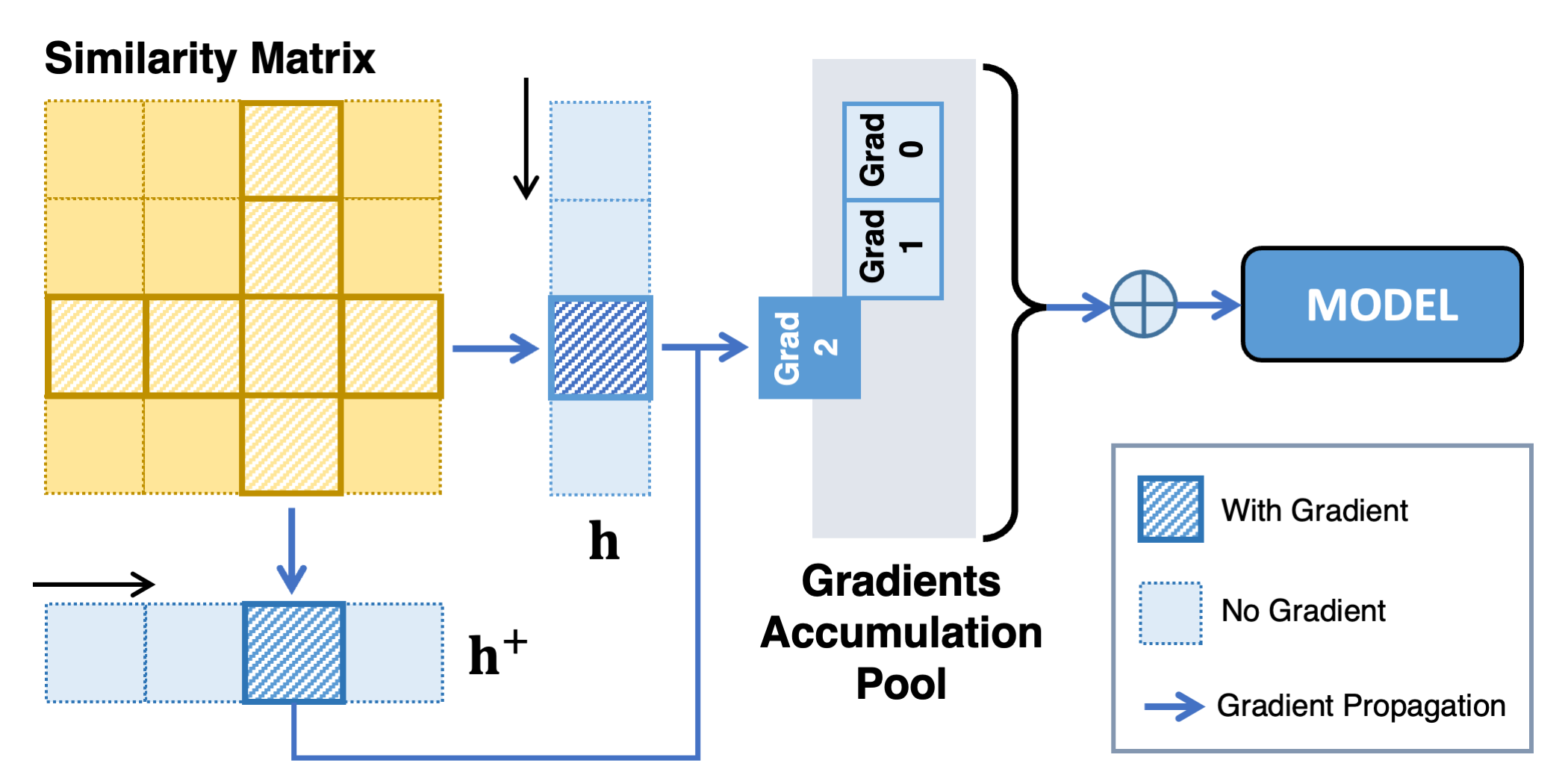}
      \end{center}
      \captionsetup{font={small}}
      \caption{Divide-and-conquer contrastive learning detail.}
      \label{figure2}
      \vspace{-0.4cm}
\end{wrapfigure}

This allows us to train the model with a larger effective batch size while preserving the memory constraints. At the same time, this method retains the advantages of end-to-end contrastive learning, including a synchronous update of the encoder and a lossless comparison of positive and negative samples. It has been mathematically rigorous and proved to be completely equivalent to the end-to-end contrastive learning method, the mathematical proof is detailed in Appendix \ref{sec:appendix1}, where we provide a demonstration of the following: 
\[ \forall\omega\in\Omega,\enspace \frac{\partial \mathcal{L}}{\partial \omega} = \sum_{k=1}^{S} \frac{\partial \mathcal{L}^{(k)}}{\partial \omega}, \]
where $\Omega$ is the parameter set of the model and $S$ is the number of steps during the divide-and-conquer contrastive learning. $\mathcal{L}$ and $\mathcal{L}^{(k)}$ is the loss of the end-to-end contrastive learning and the $k$-th step of the devide-and-conquer contrastive learning, respectively.

The diagram of divide-and-conquer contrastive learning is shown in Fig.\ref{figure2}, and its details are presented as follows:

\textbf{Step 1:} For a large batch size (denoted as $T$), we first pass all inputs through the encoder in chunks without saving gradients, computing all $h_i$ and $h_i^+$ ($1\leq i\leq T$).

\textbf{Step 2:}  Next, we partition the large batch into mini-batches (the mini-batch size is denoted as $t$) to perform encoding calculations while saving gradients. This process generates $h_j^{'}$ and $h_j^{'+}$ with gradients for each sample in the mini-batch, where $k\cdot t \leq j\leq (k+1)\cdot t$. We replace the corresponding positions of $h_j$ and $h_j^+$ with $h_j^{'}$ and $h_j^{'+}$ for the computation of the contrastive learning loss function and then perform backpropagation.

\textbf{Step 3:}  Repeat Step 2, performing computations for other mini-batches. During this process, gradient accumulation is used, and model parameters are not updated. We continue this repetition until all $T$ samples have been processed.

\subsubsection{Masked Language Modeling}
In our approach, we randomly mask the gene expression levels in the input sequence and utilize the model's output vectors at the corresponding positions to predict the reconstruction of the original input. We use the cross-entropy loss function as the loss function for this multi-classification task, defined as:
\[ \mathcal{L}_{MLM}=-\frac{1}{N}\sum_{i=1}^N\sum_{j=1}^M v_{ij}\log{(\hat{v}_{ij})}, \]
where $N$ is the number of masked genes, $M$ is the number of categories, $v_{ij}$ and $\hat{v}_{ij}$ are the label and predicted probability, respectively, for the $i$-th gene expression being assigned to the $j$-th class.

\subsubsection{Cell Type Discrimination}
Since we perform representation learning on both healthy single-cell and cancer single-cell data, we specifically incorporate a pre-training task to distinguish tumor cells from normal cells. This task aims to make the model focus on cell-level representations while also highlighting the differences between tumor cells and normal cells, thereby obtaining a better semantic understanding of both healthy and diseased cells. A \texttt{[CLS]} label is added at the beginning of each single-cell gene expression sequence, and the output at this position is used to predict whether the cell originates from tumor tissue or normal tissue. We use the cross-entropy loss function defined as:
\[ \mathcal{L}_{CLS}=-l\cdot \log{(v_{CLS})}-(1-l)\cdot \log{(1-v_{CLS})}, \]
where $l$ and $v_{CLS}$ are the label and predicted probability of this cell, respectively.

\section{Experiments}

In this section, we begin by providing an overview of the experimental settings, which encompass the datasets and baselines we chose, and the evaluation metrics employed for tasks. Subsequently, we delve into a thorough analysis of the experiment results. Due to space constraints, we focus on providing essential information and experiment details in this section, and other supplementary information (e.g. experiment conditions, details of fine-tuning, hyperparameter settings, additional ablation experiments on alternative pre-training tasks, etc.) are in Appendix \ref{sec:appendix2}, \ref{sec:appendix4}-\ref{sec:appendix6}.

\subsection{Experiment Settings}

\subsubsection{Datasets}

\paragraph{Pre-training data} During pre-training, we utilize almost 2 million scRNA-seq data, obtain from two distinct sources: PanglaoDB \cite{franzen2019panglaodb} and CancerSCEM \cite{zeng2022cancerscem}. PanglaoDB is an online database of open scRNA-seq data resources, and we leverage 1,126,580 single-cell data points from 74 human tissues in our pre-training process. CancerSCEM is a curated collection of high-quality human cancer scRNA-seq data from the literature, and we utilize 638,341 single-cell data points from 208 cancer samples. The specific datasets are chosen based on their relevance to our research on the understanding and application of normal and cancer cells.

\paragraph{Downstream tasks}
We evaluate the representation ability of CellLM on two different kinds of tasks at two cell data levels. The first task is cell type annotation which involves predicting cell types from scRNA-seq data representation and directly verifies the representation power. We use human peripheral blood mononuclear cells (PBMCs) dataset Zheng68k \cite{zheng2017massively} and pancreas dataset Baron \cite{baron2016single} for this task.

The second task is drug sensitivity prediction which requires to predict drug sensitivity through gene expression values. The single-cell level and cell line level are used in the drug sensitivity prediction task. For single-cell task, we conduct full and few-shot scenarios experiments on two datasets, human lung cancer cells (GSE149383) \cite{aissa2021single} and human oral squamous cancer cells (GSE117872) \cite{sharma2018longitudinal, ravasio2020single, suphavilai2021predicting}. For the cell line task, we evaluate the representation ability of CellLM on cell lines in drug sensitivity data integrating from CCLE \cite{barretina2012cancer} and GDSC \cite{iorio2016landscape}.

\subsubsection{Baselines and Evaluations}
\textbf{Cell type annotation.} scBERT is the first single-cell LLM and is the best-performing model on this task. To provide more comprehensive comparisons, we select Scanpy \cite{SCANPY2018, SCANPY2023}, a widely used tool for single-cell analysis, as our baseline. We choose macro $F_1$-score, weighted $F_1$-score, and accuracy as the evaluation metric.

\textbf{Drug sensitivity prediction.} We compare the capabilities of scBERT on single-cell data. In the case of cell line data, we compare our results with both scBERT and DeepCDR \cite{liu2020deepcdr}, utilizing single-omics data. Single-cell drug sensitivity prediction predicts whether the cell is sensitive to the drug, so we use $F_1$-score for evaluation. Cell line drug sensitivity prediction is to predict the IC50 of the drug experiment, which is a regression task. Therefore, we use Pearson's correlation, RMSE, MAE, and $R^2$ for evaluation. For the cell line drug sensitivity prediction task, we implement a simplified version of the drug encoder in TGDRP \cite{Zhu2021TGSAPA}.

\begin{table}[htbp]
  \captionsetup{font={small}}
  \caption{Results of cell type annotation reported in percentages (\%). $^\dag$ stands for the reproduction results, $_\mathrm{MoCo}$ stands for we change the contrastive learning module in pre-training with MoCo, and $\mathrm{_{w/o\;CL}}$ means contrastive learning is not used during pre-training.}
  \label{cell_type_annotation_table}
  \centering
  \resizebox{\textwidth}{!} {
  \begin{tabular}{lccc|ccc}
    \toprule
    \multicolumn{1}{c}{} & \multicolumn{3}{c}{\textbf{Zheng68K}}& \multicolumn{3}{c}{\textbf{Baron}} \\
    \cmidrule(r){1-4}\cmidrule(r){5-7}
    Model          & macro $F_1$  & weighted $F_1$ & Accuracy  & macro $F_1$   & weighted $F_1$  & Accuracy \\
    \midrule
    Scanpy$^\dag$ & 49.5$_{\pm{2.2}}$ & 60.6$_{\pm{2.3}}$ & 63.5$_{\pm{2.6}}$ & 61.5$_{\pm{2.2}}$ & 90.1$_{\pm{1.0}}$ & 91.7$_{\pm{1.0}}$  \\
    scBERT$^\dag$  & 68.8$_{\pm0.6}$ & 77.5$_{\pm0.8}$ & 77.9$_{\pm0.8}$ & 84.4$_{\pm4.1}$ & 97.0$_{\pm0.3}$ & 97.2$_{\pm0.3}$ \\
    \midrule  
    \specialrule{0em}{1.5pt}{1.5pt}
    \midrule
    CellLM$\mathrm{_{w/o\;CL}}$ & 66.4$_{\pm0.9}$ & 76.1$_{\pm1.1}$ & 76.6$_{\pm1.2}$ & 85.5$_{\pm3.1}$ & 97.6$_{\pm0.3}$ & 97.8$_{\pm0.3}$ \\
    CellLM$\mathrm{_{MoCo}}$  & 67.0$_{\pm1.2}$ & 75.2$_{\pm1.5}$ & 75.6$_{\pm1.6}$ & 89.0$_{\pm2.4}$ & 97.3$_{\pm0.2}$ & 97.4$_{\pm0.2}$ \\
    \midrule    
    CellLM  & \textbf{71.8}$_{\pm1.0}$ & \textbf{80.1}$_{\pm1.1}$ & \textbf{81.0}$_{\pm0.9}$ & \textbf{90.1}$_{\pm2.3}$ & \textbf{98.1}$_{\pm0.1}$ & \textbf{98.2}$_{\pm0.2}$ \\
    \bottomrule
  \end{tabular}
  }
\end{table}

\subsection{Experiment Results}

\subsubsection{Results of Cell Type Annotation} 

\begin{figure}[htbp]
    \vspace{-0.6cm}
    \captionsetup{font={small}}
    \centering
    \subfigure[Results of Zheng68K]{\includegraphics[width=0.4\linewidth]{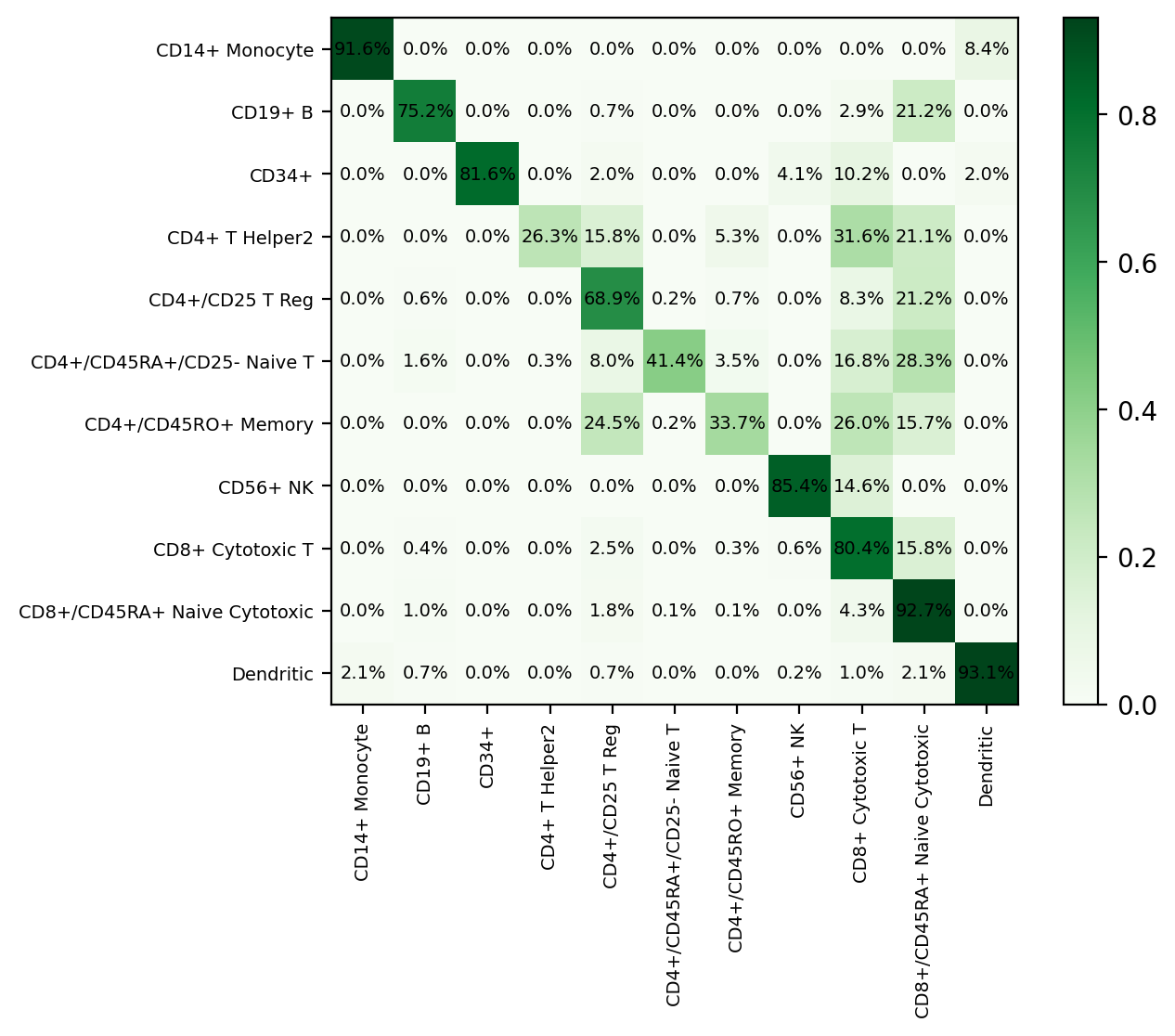}}
    \subfigure[Results of Baron]{\includegraphics[width=0.4\linewidth]{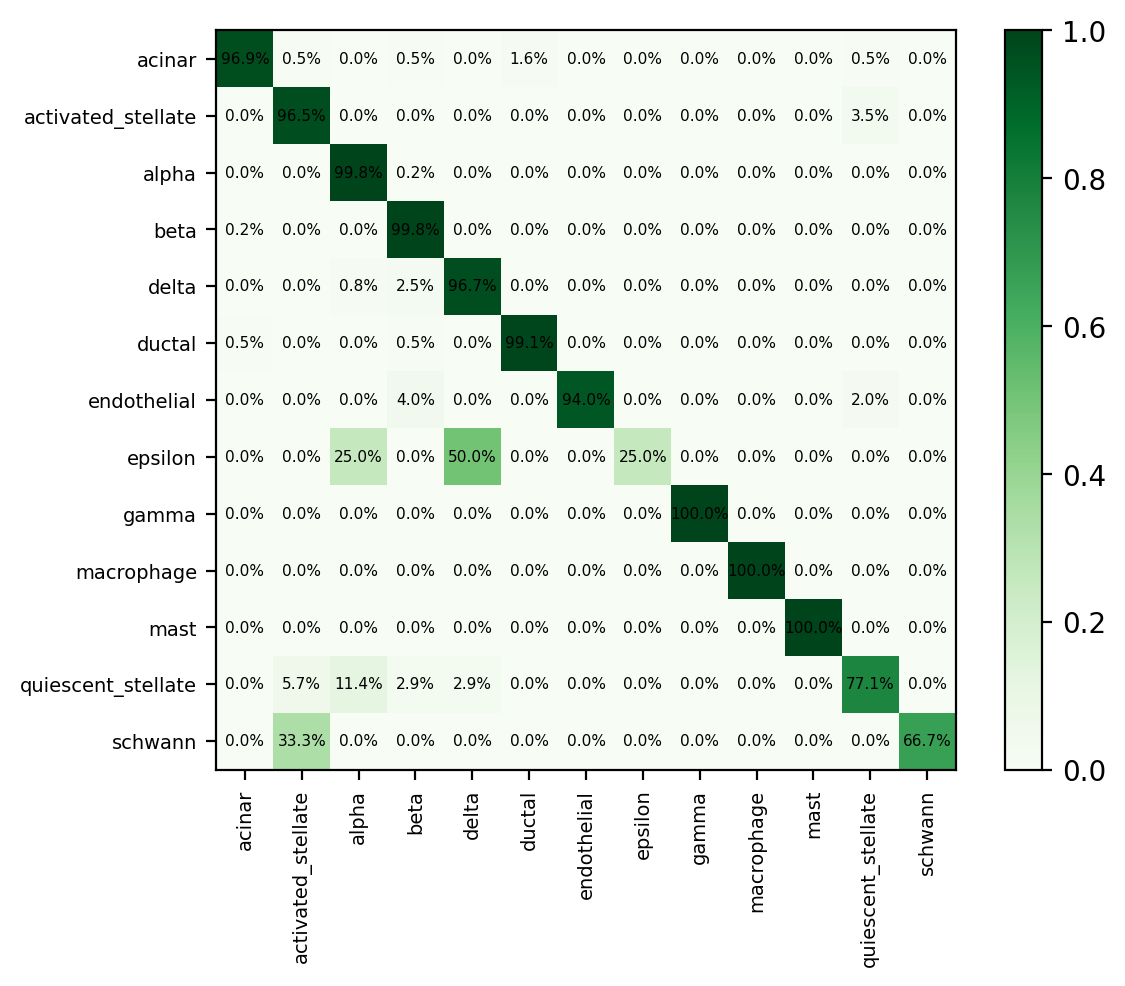}}
    \captionsetup{font={small}}
    \caption{Heatmap for the confusion matrix of results on cell type annotation.}
    \label{fig_heatmap}
\end{figure}

First, we evaluate the representation ability of CellLM on the task of single-cell type annotation. The results present in Table \ref{cell_type_annotation_table} and the heatmap of the confusion matrix in Fig.\ref{fig_heatmap} demonstrate that our model achieves state-of-the-art (SOTA) performance on Zheng68K and Baron datasets. Notably, the macro $F_1$-score surpasses the current leading model scBERT by 3.0\% and 5.7\%, respectively. Our method successfully optimizes the original feature space of BERT and significantly improves the representation degradation caused by anisotropy in the embedding space. Moreover, these findings validate the effectiveness of the divide-and-conquer contrastive learning approach.

\subsubsection{Results of Drug Sensitivity Tasks}

In the single-cell drug sensitivity prediction experiments, we conduct both few-shot and full data scenarios. In the few-shot experiment, we aim to simulate the scarcity of drug sensitivity data, which closely resembles real-world scenarios. Here, we only utilize 5\% of the available data samples for training. Conversely, the full data experiment employs 80\% of the data for training. 

\begin{figure}[htb]
    \centering
	\begin{minipage}[b]{0.5\linewidth}
		\centering
            \resizebox{\textwidth}{!}{
            \small
            \begin{tabular}{lcc} 
            \toprule
            \multicolumn{3}{c}{\textbf{GSE149383}}          \\
            \cmidrule(r){1-3}
            Model     & Few-Shot  $F_1$-score (\%) &  Full Data  $F_1$-score (\%) \\
            \midrule
            scBERT$^\ddag$  & 73.8$_{\pm8.5}$ & 97.1$_{\pm{6.9}}$ \\
            CellLM  & \textbf{88.7}$_{\pm1.7}$ & \textbf{98.4}$_{\pm{1.7}}$ \\
            \midrule
            \specialrule{0em}{1.5pt}{1.5pt}
            \midrule
            \multicolumn{3}{c}{\textbf{GSE117872}}\\
            \cmidrule(r){1-3}
            Model     & Few-Shot  $F_1$-score (\%)  &  Full Data  $F_1$-score (\%) \\
            \midrule
             scBERT$^\ddag$  & 87.4$_{\pm4.3}$ & 97.2$_{\pm{3.6}}$ \\
            CellLM   & \textbf{89.0}$_{\pm1.0}$ & \textbf{99.2}$_{\pm{1.1}}$ \\
            \bottomrule
            \end{tabular}}
           \captionsetup{font={small}}
           \captionof{table}{Results of single-cell drug sensitivity in few-shot and full data scenario. $^\ddag$  indicates that we use scBERT as the cell encoder.\label{sc_DS_table}}
	    \end{minipage}
            \hfill
    \begin{minipage}[b]{0.45\linewidth}
		\centering
		\makebox[\textwidth][c]    {\includegraphics[width=0.8\linewidth]{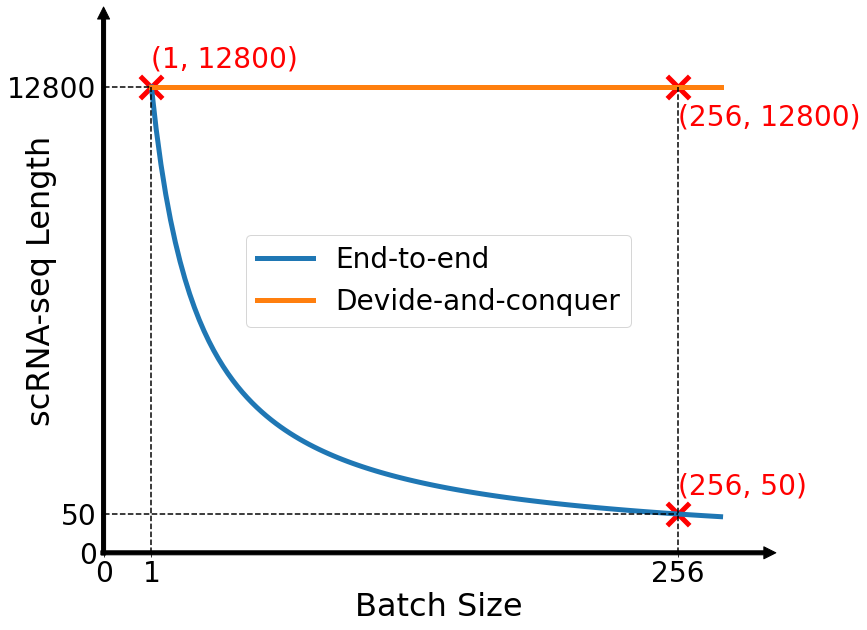}}
            \captionsetup{font={small}}
            \caption{The relationship between the max scRNA-seq length and the batch size for contrastive learning under a 40 GB GPU memory.}
            \label{figure4}
    \end{minipage}
\end{figure}

The experiment results are summarized in Table \ref{sc_DS_table}. Comparing with scBERT, CellLM demonstrates substantial improvements in both scenarios across the two datasets. Specifically, in the few-shot scenario, CellLM achieves an average improvement of 8.25\%. In the full-data scenario, the average improvement is 1.65\%. These experiment outcomes highlight the enhanced understanding of cancer obtained by CellLM, as it surpasses scBERT, which is solely pre-trained on normal human tissues. The incorporation of cancer data enables CellLM to capture cancer-specific features, leading to superior representations and significantly improved performance in cancer drug virtual screening.

Table \ref{cell_line_DS_table} presents the results of the drug sensitivity prediction experiment conducted on cell lines. We simulate two scenarios that resemble real-world drug repurposing efforts \cite{Ye2021AUD}: cell line warm start and cell line cold start. In the warm-start scenario, the data is randomly divided into training, validation, and test sets. Conversely, in the cold-start scenario, the test set contains cell lines that are not seen during training. This setup requires the model to predict drug responses for novel cell lines it has never encountered before. Since our focus is on cell embedding rather than drug-specific performance, we do not establish a specific experiment scenario for drug cold start.

\begin{table}[htbp]
    \small
    \centering
    \captionsetup{font={small}}
    \caption{Results of single-omics cell line drug sensitivity in warm and cold start. $^\dag$ stands for the reproduction results on single-omics data, and $^\ddag$ indicates that we use scBERT as the cell encoder.}
    \label{cell_line_DS_table}
    \resizebox{\textwidth}{!}{
    \begin{tabular}{lcccc|cccc}
        \toprule
        \multicolumn{1}{c}{} & \multicolumn{4}{c}{\textbf{Warm Start}}  & \multicolumn{4}{c}{\textbf{Cold Start}}  \\
        \cmidrule(r){1-5}\cmidrule(r){6-9}
        Model     & {\makecell[c]{Pearson's \\ correlation}} & $R^2$ & RMSE & MAE  & {\makecell[c]{Pearson's \\ correlation}} & $R^2$  & RMSE & MAE  \\
        \midrule
        DeepCDR$^\dag$ & 0.863$_{\pm{0.005}}$  & 0.742$_{\pm{0.010}}$ & 1.392$_{\pm{0.026}}$ & 1.053$_{\pm{0.024}}$ & 0.847$_{\pm{0.009}}$ & 0.716$_{\pm{0.016}}$ & 1.447$_{\pm{0.041}}$ & 1.083$_{\pm{0.030}}$  \\
        scBERT$^\ddag$ & 0.872$_{\pm0.012}$ & 0.758$_{\pm0.023}$ & 1.334$_{\pm0.069}$ & 1.020$_{\pm0.055}$ & 0.857$_{\pm0.005}$ & 0.723$_{\pm0.016}$ & 1.446$_{\pm0.041}$ & 1.086$_{\pm0.036}$  \\
        CellLM  & \textbf{0.934}$_{\pm0.006}$& \textbf{0.831}$_{\pm0.013}$ & \textbf{1.129}$_{\pm0.041}$ & \textbf{0.864}$_{\pm0.031}$  & \textbf{0.885}$_{\pm0.006}$ & \textbf{0.779}$_{\pm0.012}$  & \textbf{1.261}$_{\pm0.033}$ & \textbf{0.942}$_{\pm0.037}$  \\
        \bottomrule
    \end{tabular}}
\end{table}

Compared with DeepCDR, CellLM demonstrates absolute improvements of 6.2\% and 3.8\% in Pearson's correlation coefficient under the two scenarios. The experiment results indicate that the knowledge acquired by CellLM from single-cell transcriptomics remains effective when transferred to cell lines. Furthermore, CellLM's performance exceeds that of current methods for single-omics data.

\subsubsection{Ablation Study of Divide-and-Conquer Contrastive Learning}
To demonstrate the effectiveness of the divide-and-conquer contrastive learning method, we conduct two ablation experiments. The first one is pre-training without contrastive learning, and the second one replaces the divide-and-conquer contrastive learning with MoCo. Their performance on the cell annotation task is shown in Table \ref{cell_type_annotation_table} (CellLM$\mathrm{_{w/o\;CL}}$ and CellLM$\mathrm{_{MoCo}}$). 
Compared with the method that does not utilize contrastive learning (CellLM$\mathrm{_{w/o\;CL}}$), CellLM demonstrates a substantial absolute improvement in macro $F_1$-score, reaching 5.4\% and 4.6\%. Furthermore, when compared to the method employing MoCo (CellLM$\mathrm{_{MoCo}}$), CellLM achieves an absolute improvement of 4.8\% and 1.1\% in macro $F_1$-score, respectively. The results indicate that, firstly, the addition of contrastive learning can indeed improve cell representation learning using only a BERT-based structure. Secondly, the lossless end-to-end contrastive learning method, divide-and-conquer contrastive learning, can not only achieve a large training batch size but also solve the performance degradation caused by the asynchronous update of positive and negative sample encoders.

Ablation experiments on other self-supervised tasks in pre-training are presented in Appendix \ref{sec:appendix2}. In addition, we also explore the interpretability of CellLM, which can be found in Appendix \ref{sec:appendix3}.

\subsection{Analysis: The Limit of Computing under Fixed GPU Memory}
Due to the utilization of Performer-based cell encoders, the GPU memory requirements during training are roughly proportional to the length of the scRNA-seq data and the mini-batch size. Therefore, with the fact of a fixed GPU memory limitation, the max scRNA-seq length and mini-batch size that can be employed are approximately inversely proportional. 

In our work, we use NVIDIA Tesla A100 40G and set the batch size for contrastive learning to 256 to train the model. If we use the end-to-end contrastive learning method, the max scRNA-seq length would only be 50, which is clearly insufficient since the number of expressed genes in most single-cell data ranges from 300 to 5000. In contrast, with the divide-and-conquer contrastive learning approach, we can split any batch size into mini-batches with a minimum length of 1. Under the aforementioned experiment settings, the max gene sequence length can be set to approximately 13,000, which effectively covers the majority of single-cell and cell line data, enabling the model to learn cell representations more effectively. Fig.\ref{figure4} provides a schematic illustration of this analysis.

\section{Conclusion}

In this paper, we propose the divide-and-conquer contrastive learning method to decouple the batch size from the GPU memory size so as to solve the problem that the batch size is limited by GPU memory size when using contrastive learning to optimize cell representation. This approach allows training with arbitrarily large batch sizes on limited memory, which provides more flexible and extensive choices for modeling scRNA-seq data. In addition, we design self-supervised tasks to help the model better learn to understand features between cancer and normal cell. The SOTA results of CellLM in a range of downstream tasks confirm its effectiveness on cell representation. Furthermore, we firmly believe that the utility of divide-and-conquer contrastive learning extends far beyond its application in the cell representation scenario. Its potential can be harnessed to tackle similar challenges encountered in various other domains and contexts.

\section{Limitations}

We focus solely on utilizing scRNA-seq data, yet there exist additional omics data such as scATAC-seq that can provide valuable insights. Furthermore, the current experiment exclusively focuses on normal and cancer cells, overlooking other cell types and diseases. However, we firmly believe that broadening the range of data types will significantly enhance the model's applicability across various downstream tasks. With this vision in mind, our future research endeavors will center around investigating the representation of multi-omics data and incorporating diverse cell types, enabling us to unlock new insights and possibilities.

\section*{Acknowledgements}

This work is jointly supported by the National Key R\&D Program of China (No. 2022YFF1203002). 

\newpage

\small
\bibliographystyle{unsrt}
\bibliography{ref_main}

\newpage
\setcounter{section}{0}
\setcounter{equation}{0}
\setcounter{subsection}{0}
\renewcommand{\theequation}{A.\arabic{equation}}
\renewcommand{\thesubsection}{\Alph{subsection}}
\renewcommand{\thetable}{\Alph{table}}
\renewcommand{\thefigure}{\Alph{figure}}
\numberwithin{figure}{subsection}
\numberwithin{table}{subsection}

\section*{Appendix}

\subsection{Proof of gradient equivalence}
\label{sec:appendix1}

Here we prove mathematically that the model parameter gradients obtained by this computation method are exactly the same as those obtained by direct end-to-end contrastive learning. 

Let the input vectors and their augmented version be denoted as $\bm{x}$ and $\bm{x}^+$ respectively, where $\bm{x},\bm{x}^+\in \mathbb{R}^{T\times d_{in}}$, $T$ is the batch size, and $d_{in}$ is the input dimension. The model is denoted as $f$, with its parameter set represented by $\Omega$. Thus, we have:
\begin{gather*}
    \bm{h} = f(\bm{x}) \\
    \bm{h}^+ = f(\bm{x}^+) \\
\end{gather*}
Next, we use the model outputs $\bm{h},\bm{h}^+\in \mathbb{R}^{T\times d_{out}}$ to calculate the similarity matrix $\bm{M}=\{m_{ij}\}$, and compute the cross entropy loss $\mathcal{L}$ by comparing it with the identity matrix $I$. The specifics are as follows:
\begin{gather*}
    \forall m_{ij} \in \bm{M}, \; m_{ij} = \frac{\bm{h}_i\cdot \bm{h}_j^+}{\Vert \bm{h}_i \Vert_2 \cdot \Vert \bm{h}_j^+ \Vert_2} \\
    \mathcal{L} = \mathcal{L}(\bm{M}, \bm{I})
\end{gather*}

In \textbf{end-to-end} contrastive learning, the gradient computation process for each model parameter during backpropagation is as follows:
\[ \forall\omega\in\Omega,\enspace \omega.grad = \frac{\partial \mathcal{L}}{\partial \omega} = \frac{\partial \mathcal{L}}{\partial \bm{h}} \cdot \frac{\partial \bm{h}}{\partial \omega} + \frac{\partial \mathcal{L}}{\partial \bm{h}^+} \cdot \frac{\partial \bm{h}^+}{\partial \omega} \]

In our \textbf{devide-and-conquer} contrastive learning, we divide the large batch into $S$ small batches with a length of $t$ each, and the gradient computation is performed in $S$ steps and accumulated.  In the $k$-th step, only a small part of $\bm{h}, \bm{h}^+$ require gradients, denoted as $\bm{h}^{(k)}, \bm{h}^{+(k)} \in \mathbb{R}^{t\times d_{out}}$. That is, $\bm{h}, \bm{h}^+$ are divided into:
\[ \bm{h} = \begin{bmatrix} \bm{h}^{(0)} \\ \bm{h}^{(1)} \\ \vdots \\ \bm{h}^{(S-1)} \end{bmatrix},\; \bm{h}^+ = \begin{bmatrix} \bm{h}^{+(0)} \\ \bm{h}^{+(1)} \\ \vdots \\ \bm{h}^{+(S-1)} \end{bmatrix} \]
The resulting gradient for $\omega$ in the $k$-th step is denoted as $\omega.grad^{(k)}$. Then, the gradient obtained in the $k$-th step is given by:
\[ \omega.grad^{(k)} = \left(\frac{\partial \mathcal{L}}{\partial \omega}\right)^{(k)} = \frac{\partial \mathcal{L}}{\partial \bm{h}^{(k)}} \cdot \frac{\partial \bm{h}^{(k)}}{\partial \omega} + \frac{\partial \mathcal{L}}{\partial \bm{h}^{+(k)}} \cdot \frac{\partial \bm{h}^{+(k)}}{\partial \omega} \]

Using gradient accumulation, the total gradient obtained is given by $\sum_{k=1}^{S}{\omega.grad^{(k)}}$. Below, we will prove that it is the same as the gradient obtained by end-to-end training:
\begin{align*}
    \because\qquad & \bm{h} = \begin{bmatrix} \bm{h}^{(0)} \\ \bm{h}^{(1)} \\ \vdots \\ \bm{h}^{(S-1)} \end{bmatrix} \\
    \therefore\qquad & \frac{\partial \mathcal{L}}{\partial \bm{h}} = \begin{bmatrix} \partial \mathcal{L} / \partial \bm{h}^{(0)} \\ \partial \mathcal{L} / \partial \bm{h}^{(1)} \\ \vdots \\ \partial \mathcal{L} / \partial \bm{h}^{(S-1)} \end{bmatrix} , \;
    \frac{\partial \bm{h}}{\partial \omega} = \begin{bmatrix} \partial \bm{h}^{(0)} / \partial \omega \\ \partial \bm{h}^{(1)} / \partial \omega \\ \vdots \\ \partial \bm{h}^{(S-1)} / \partial \omega \end{bmatrix} \\
    \therefore\qquad & \frac{\partial \mathcal{L}}{\partial \bm{h}} \cdot \frac{\partial \bm{h}}{\partial \omega} = \sum_{k=1}^{S}(\frac{\partial \mathcal{L}}{\partial \bm{h}^{(k)}} \cdot \frac{\partial \bm{h}^{(k)}}{\partial \omega}) \\
    &Similarly,\quad  \frac{\partial \mathcal{L}}{\partial \bm{h}^+} \cdot \frac{\partial \bm{h}^+}{\partial \omega} = \sum_{k=1}^{S}(\frac{\partial \mathcal{L}}{\partial \bm{h}^{+(k)}} \cdot \frac{\partial \bm{h}^{+(k)}}{\partial \omega}) \\
    \therefore\qquad & \omega.grad = \sum_{k=1}^{S}(\frac{\partial \mathcal{L}}{\partial \bm{h}^{(k)}} \cdot \frac{\partial \bm{h}^{(k)}}{\partial \omega} + \frac{\partial \mathcal{L}}{\partial \bm{h}^{+(k)}} \cdot \frac{\partial \bm{h}^{+(k)}}{\partial \omega}) = \sum_{k=1}^{S}{\omega.grad^{(k)}} \\
   & \bm{Q.E.D.}
\end{align*}

\subsection{Additional Ablation Study}
\label{sec:appendix2}
In the main paper, we perform ablation experiments related to contrastive learning. Here we would like to perform two additional ablation experiments, one to illustrate the benefit of introducing cancer cells in our pre-training and adding the pre-training task of distinguishing normal cells from cancer cells; the other to demonstrate the boost from larger scale models and to illustrate the scalability of our model.

\begin{itemize}
\item To demonstrate the effectiveness of the task of distinguishing between cancer and normal cells, we perform pre-training with this task removed. We test its performance on the cell line drug sensitivity prediction task, as shown in Table \ref{cell_line_DS_table_app} (CellLM$\mathrm{_{w/o\;CLS}}$). Compared to the method without the cell classification task, CellLM demonstrates an absolute improvement of 1.4\% and 1.5\% in Pearson's correlation coefficients for cell line warm-start and cold-start, respectively, reflecting the improvement brought by our inclusion of the diseased cell pre-training task.

\item To test the performance improvement brought by the larger-scale model, we train two simplified versions of CellLM with reduced parameters (CellLM$_{25M}$ and CellLM$_{8M}$) and evaluate their performance on the cell type annotation task. The results in Table \ref{cell_type_annotation_table_app} show that the performance of CellLM improved with the increase in model size, which reflects the scalability of the model and provides the possibility of using more data to train industrial-level large models in the future.
\end{itemize}

\begin{table}[htpb]
    \small
    \centering
    \caption{Results of single-omics cell line drug sensitivity in warm and cold start. }
    \label{cell_line_DS_table_app}
    \resizebox{\textwidth}{!}{
    \begin{tabular}{lcccc|cccc}
        \toprule
        \multicolumn{1}{c}{} & \multicolumn{4}{c}{\textbf{Warm Start}}  & \multicolumn{4}{c}{\textbf{Cold Start}}  \\
        \cmidrule(r){1-5}\cmidrule(r){6-9}
        Model     & {\makecell[c]{Pearson's \\ correlation}} & $R^2$ & RMSE & MAE  & {\makecell[c]{Pearson's \\ correlation}} & $R^2$  & RMSE & MAE  \\
        \midrule
        CellLM$\mathrm{_{w/o\; CLS}}$ & 0.898$_{\pm 0.010}$ & 0.804$_{\pm 0.019}$ & 1.216$_{\pm0.055}$ & 0.900$_{\pm0.052}$ & 0.870$_{\pm0.006}$ & 0.756$_{\pm0.017}$ & 1.344$_{\pm0.044}$ & 1.029$_{\pm0.035}$ \\
        CellLM  & \textbf{0.912}$_{\pm0.006}$& \textbf{0.831}$_{\pm0.013}$ & \textbf{1.129}$_{\pm0.041}$ & \textbf{0.864}$_{\pm0.031}$  & \textbf{0.885}$_{\pm0.006}$ & \textbf{0.779}$_{\pm0.012}$  & \textbf{1.261}$_{\pm0.033}$ & \textbf{0.942}$_{\pm0.037}$  \\
        \bottomrule
    \end{tabular}}
\end{table}

\begin{table}[htpb]
  \caption{Results of cell type annotation reported in percentages (\%). }
  \label{cell_type_annotation_table_app}
  \centering
  \resizebox{\textwidth}{!} {
  \begin{tabular}{lccc|ccc}
    \toprule
    \multicolumn{1}{c}{} & \multicolumn{3}{c}{\textbf{Zheng68K}}& \multicolumn{3}{c}{\textbf{Baron}} \\
    \cmidrule(r){1-4}\cmidrule(r){5-7}
    Model          & macro $F_1$  & weighted
    $F_1$ & Accuracy  & macro $F_1$   & weighted $F_1$  & Accuracy \\
    \midrule
    CellLM$\mathrm{_{8M}}$ & 68.0$_{\pm1.1}$ & 77.4$_{\pm1.3}$ & 77.6$_{\pm1.4}$ & 86.7$_{\pm2.7}$ & 97.6$_{\pm0.2}$ & 97.7$_{\pm0.3}$ \\
    CellLM$\mathrm{_{25M}}$ & 70.4$_{\pm0.6}$ & 78.3$_{\pm0.8}$ & 79.3$_{\pm1.2}$ & 89.1$_{\pm1.9}$ & 97.7$_{\pm0.1}$ & 97.9$_{\pm0.1}$ \\
    \midrule    
    CellLM$\mathrm{_{50M}}$  & \textbf{71.8}$_{\pm1.0}$ & \textbf{80.1}$_{\pm1.1}$ & \textbf{81.0}$_{\pm0.9}$ & \textbf{90.1}$_{\pm2.3}$ & \textbf{98.1}$_{\pm0.1}$ & \textbf{98.2}$_{\pm0.2}$ \\
    \bottomrule
  \end{tabular}
  }
\end{table}

\subsection{Interpretability Study}
\label{sec:appendix3}
Interpretability study is an important issue in deep learning. CellLM encodes each gene separately in scRNA-seq data, preserving a high level of interpretability at the gene level. Additionally, since CellLM uses the generalized attention mechanism from Performer during its inference process, the degree of attention paid to each gene during inference is visible.

\begin{figure}[htp]
  \centering
  \makebox[0.8\textwidth][c]{\includegraphics[width=1\linewidth]{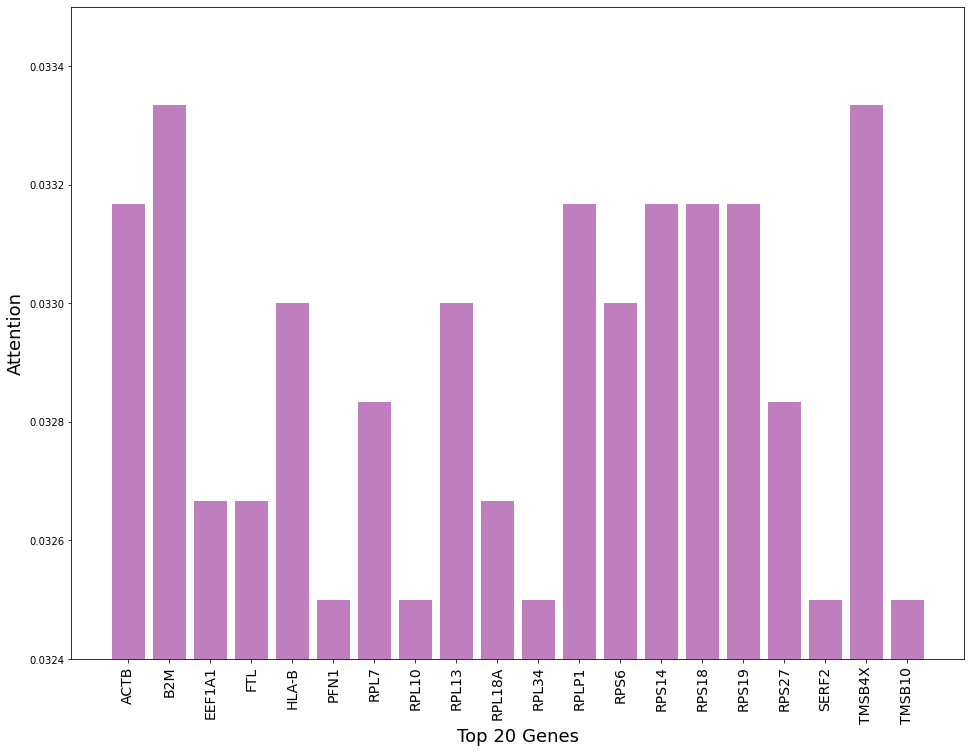}}
  \caption{The attention levels of the top attention genes when the model distinguishes between normal cells and cancer cells.}
  \label{figure1_app}
\end{figure}

\begin{figure}[htbp]
    \vspace{-0.6cm}
    \centering
    \subfigure[Results of the top 20 expressed genes]{\includegraphics[width=0.6\linewidth]{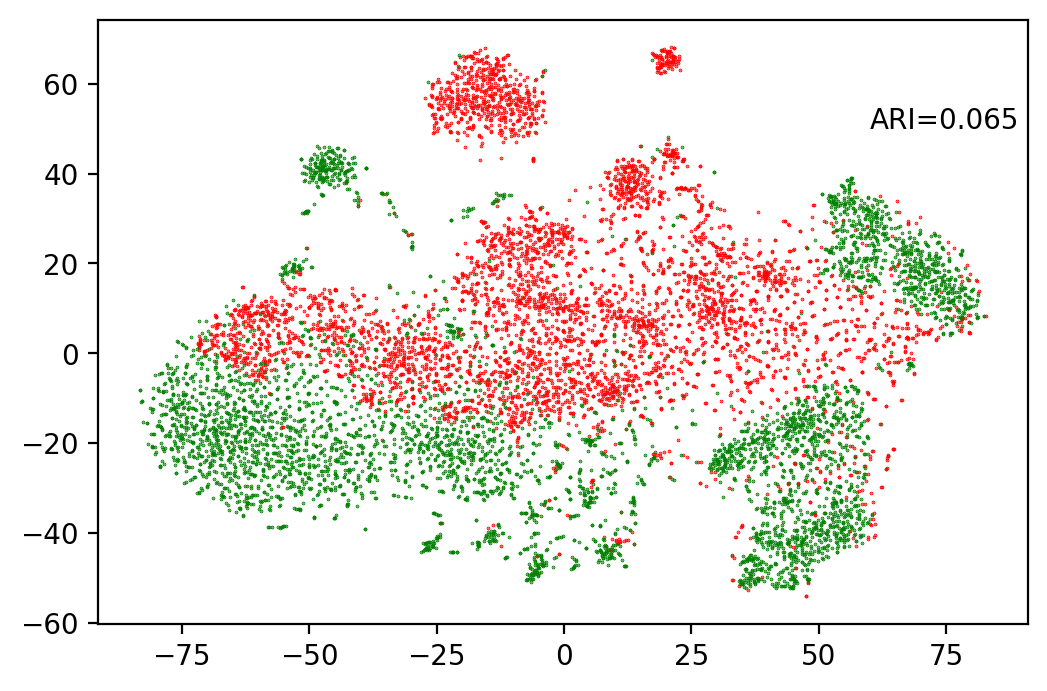}}
    \subfigure[Results of the top 20 genes with the highest attention]{\includegraphics[width=0.6\linewidth]{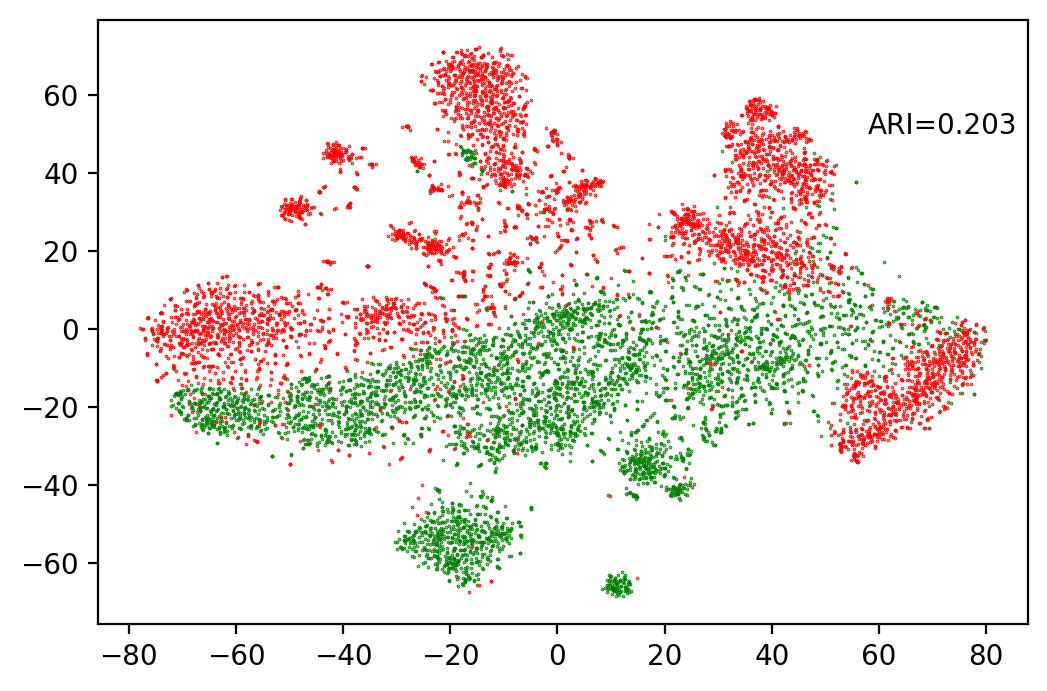}}
    \captionsetup{font={small}}
    \caption{A t-SNE representation of cells using the expression levels of only 20 genes, coloured by the cell types (normal and cancer).}
    \label{fig_cluster}
    \vspace{-0.3cm} 
\end{figure}

In the pre-training task of distinguishing between cancer cells and normal cells, we use the \texttt{[CLS]} token for cell classification. By observing the attention maps generated during model inference, we can see the degree of attention paid to each gene by the model during classification. Fig.\ref{figure1_app} shows the top 20 genes that receive the most attention. Genes that are highly attended to when differentiating between cancer and normal cells likely reflect differences in gene expression levels between these two cell types. To illustrate this point, we select 20 key genes from each of 5,000 normal cells and 5,000 cancer cells from PanglaoDB \citeapp{panglaodb} and CancerSCEM \citeapp{cancerscem}, respectively, and perform two sets of clustering. The first set simply uses the top 20 expressed genes across all 10,000 cells, while the second set uses the top 20 genes with the highest attention. As depicted in Fig.\ref{fig_cluster}, when colored by cell type, the second clustering set exhibits better separation between the two cell types (ARI: 0.203 versus 0.065), demonstrating that the genes receiving the highest attention more accurately reflect the differences between normal and cancer cells.

Moreover, the three genes with the highest degree of attention are found to be B2M (beta-2-microglobulin), TMSB4X (thymosin beta 4 X-linked), and ACTB (actin beta), which have been shown in several biological studies to be closely associated with cancer, as shown in Table \ref{cite_info}. In addition, the gene B2M, which has the highest attention level, is a well-known tumor suppressor gene \citeapp{castro2019elevated} that has been widely used for cancer detection in clinical settings \citeapp{medlineplus}. This also demonstrates the potential of CellLM to discover marker genes for specific problems.

\begin{table}[htbp]
    \caption{Evidence closely linking the three genes with the most attention from CellLM to cancer.}
    \label{cite_info}
    \centering
    \resizebox{0.9\textwidth}{!} {
        \begin{tabular}{lc}
        \toprule
        \textbf{Genes} & \textbf{Key Descriptions and References } \\
        \midrule
        B2M & {\makecell[l]{ \textbf{-} B2M mutations have been identified in considerable cancer patients. \citeapp{WANG202196} \\ \textbf{-} Immune evasion is one of the hallmark traits characteristic of cancer cells \\ and somatic B2M mutation is a mechanism of immune evasion. \citeapp{castro2019elevated} }} \\
        \midrule
        TMSB4X  & {\makecell[l]{\textbf{-} TMSB4X has been shown to be upregulated in a wide variety of human \\ carcinomas and has been implicated to be involved in altering the motility \\ of  certain tumors.  \citeapp{wang2004overexpression} \\ \textbf{-} TMSB4X is one of the most significant candidate biomarkers of some \\ types of cancer. \citeapp{chi2017global}}} \\
        \midrule
        ACTB & {\makecell[l]{\textbf{-} ACTB is abnormally expressed in multiple cancers and hence changes \\ the cytoskeleton to affect the invasiveness and metastasis of tumors. \citeapp{gu2021pan} \\ \textbf{-} Accumulating evidence indicates that ACTB is de-regulated in the liver,\\ melanoma, renal, colorectal, gastric, pancreatic, esophageal, lung, breast,\\ prostate, ovarian cancers, leukemia, and lymphoma. \citeapp{guo2013actb}}}  \\
        \bottomrule
        \end{tabular}
    }
\end{table}

\subsection{Datasets}
\label{sec:appendix4}

\subsubsection{Cell Type Annotation.} 
The annotation of single-cell types based on RNA gene expression profiles is a crucial application of scRNA-seq data because researchers typically do not know the specific type of cell being sequenced during the sequencing stage. The automatic annotation of cell types based on deep learning approaches could bring significant benefits to researchers in terms of convenience and accuracy. 

Zheng68k \citeapp{zheng68k} a highly relevant and challenging dataset, consisting of 68,450 human peripheral blood mononuclear cells (PBMCs) with 11 highly related cell types. Zheng68k provides high-quality cell type annotations, making it an ideal benchmark for evaluating annotation approaches. However, the dataset poses significant challenges due to the large number of cell categories and the uneven distribution of samples between types. Table \ref{zheng68k_info} illustrates the categories included in Zheng68K and information on the number and proportion of each type.

\begin{table}[htpb]
    \caption{The distribution of each category in Zheng68K dataset.}
    \centering
    \resizebox{0.9\textwidth}{!} {
        \begin{tabular}{lrr}
        \toprule
        \textbf{Cell type} & \textbf{\#Number} & \textbf{\#Ratio} \\
        \midrule
        CD8+ Cytotoxic T & 20,757 & 30.32\% \\
        CD8+/CD45RA+ Naive Cytotoxic & 16,645 & 24.32\% \\
        CD56+ NK & 8,775 & 12.82\%  \\
        CD4+/CD25 T Reg & 6,185 &  9.04\% \\
        CD19+ B & 5,877 &  8.59\% \\
        CD4+/CD45RO+ Memory  & 3,059 & 4.47\% \\        
        CD14+ Monocyte  & 2,847 & 4.16\% \\
        Dendritic  & 2,095 &  3.06\%\\
        CD4+/CD45RA+/CD25- Naive T  & 1,871 &  2.73\% \\
        CD34+   & 242 &  0.35\% \\
        CD4+ T Helper2  & 97 &  0.14\% \\
        \midrule
        \textsc{SUM} & 11 classes with a total of 68,450 cells. \\
        \bottomrule
        \end{tabular}
    }
    \label{zheng68k_info}
\end{table}

The original Baron dataset \citeapp{baron} comprises droplet-based single-cell RNA sequencing (scRNA-seq) data obtained from over 12,000 individual pancreatic cells. These cells are derived from four human donors and two strains of mice. In our experiment, we specifically focus on the cells relevant to humans. Due to the limited number of human T cells in the dataset (only seven), we exclude T cells from the cell-type annotation task. As a result, the final dataset used in our cell type annotation task consists of 8,562 cells categorized into 13 different cell types. Table \ref{Baron_info} illustrates the categories included in Baron and information on the number and proportion of each type.

\begin{table}[ht]
    \caption{The distribution of each category in Baron dataset.}
    \centering
    \resizebox{0.75\textwidth}{!} {
        \begin{tabular}{lrr}
        \toprule
        \textbf{Cell type} & \textbf{\#Number} & \textbf{\#Ratio} \\
        \midrule
        Beta & 2,525 &  29.47\%\\
        Alpha & 2,326 & 27.14\% \\
        Ductal & 1,077 &  12.57\% \\
        Acinar & 958 & 11.18\%  \\
        Delta & 601 &  7.01\%\\
        Activated Stellate  & 284 &  3.31\% \\        
        Gamma  & 255 &  2.98\% \\
        Endothelial  & 252 & 2.94\%  \\
        Quiescent Stellate   & 173 & 2.02\%  \\
        Macrophage   & 55 & 0.64\%  \\
        Mast  & 25 &  0.29\% \\
        Epsilon  & 18 &  0.21\% \\
        Schwann  & 13 &  0.15\% \\
        \midrule
        \textsc{SUM} & 13 classes with a total of 8,562 cells. \\
        \bottomrule
        \end{tabular}
    }
    \label{Baron_info}
\end{table}

\subsubsection{Drug Sensitivity Prediction.} Drug sensitivity prediction is a critical task in drug virtual screening, which involves using computational methods to evaluate the effectiveness of drugs. In complex diseases such as cancer, abnormal cells are often hidden among normal cells, and the lack of single-cell analysis can lead to low efficiency and high recurrence rates in drug therapy. Improving the representation of cells can improve the accuracy of drug sensitivity prediction. Cell line data which consists of many cells is often closer to the real treatment scenario than single-cell data. Therefore, we conduct drug sensitivity prediction experiments on two data levels in vitro: single-cell and cell line.

\paragraph{Single-cell drug sensitivity prediction.} At this stage, there is only experimental data available for a single drug on single-cell drug sensitivity due to various experimental difficulties and other factors. We conduct full data experiments on two datasets, human lung cancer cells (GSE149383) and human oral cancer cells (GSE117872). In addition, we also simulate few-shot scenarios to test whether the learned information in CellLM could help improve predictions when cell drug response data are scarce. Table \ref{scdr_datainfo} shows the information of data used in single-cell drug sensitivity prediction.

\begin{table}[ht]
    \caption{The data information summary of single-cell drug sensitivity prediction.}
    \label{scdr_datainfo}
    \centering
    \resizebox{1\textwidth}{!} {
        \begin{tabular}{lccccc}
        \toprule
        \textbf{GEO Accession} & \textbf{Cancer Cell Type} & \textbf{\#Cell Number} &\textbf{Drug} & \textbf{Species} &\textbf{References} \\
        \midrule
        GSE149383 & Lung cancer &  2,739 & Erlotinib & Homo Sapiens & \citeapp{aissa2021singlecell}\\
        GSE117872 & Oral squamous cell carcinomas & 1,302 & Cisplatin & Homo Sapiens &  \citeapp{sharma2018, ravasio2020singlecell, suphavilai2021} \\
        \bottomrule
        \end{tabular}
    }
\end{table}

\paragraph{Cell line drug sensitivity prediction.} The Cancer Cell Line Encyclopedia (CCLE) \citeapp{ccle} provides gene expression data of thousand of human cancer cell lines, and the Cancer Drug Sensitivity Genomics (GDSC) \citeapp{gdsc} provides experiment results of different drugs on human cancer cell lines (the most commonly used label is half-inhibitory concentration IC50, reflecting the ability of the drug to induce apoptosis). We test the transfer representation ability of CellLM from single-cell to cell lines on 106,405 pairs of drug sensitivity data, which consist of 555 cell lines and 223 drugs by integrating cell line gene expression data from CCLE and GDSC drug sensitivity data.

\subsection{Evaluation}
\label{sec:appendix5}

To provide a comprehensive overview, we will now outline the evaluation metrics employed for each task.
\paragraph{Cell type annotation.} For the cell type annotation task, we evaluate CellLM's performance using two datasets, Zheng68K and Baron, consisting of 11 classification and 13 classification tasks, respectively. 

To estimate the effectiveness of CellLM for multi-classification tasks, we employ three evaluation metrics: accuracy, macro 
$F_1$-score, and weighted $F_1$-score. Accuracy measures the closeness of the prediction to the ground truth, while macro $F_1$-score comprehensively assesses classification results without considering the importance of different categories. We also use weighted $F_1$-score to measure classification performance while accounting for the importance of different categories. These metrics are calculated based on true positive (TP), true negative (TN), false positive (FP), and false negative (FN) rates.

\begin{gather*}
    Accuracy = \frac{TP+TN}{TP+TN+FP+FN} \\
\end{gather*}
To calculate both macro $F_1$-score and weighted $F_1$-score, we need to compute Precision and Recall. These two key metrics are calculated using the following formulas:
\begin{gather*}
    Precision = \frac{TP}{TP+FP}, \quad Recall = \frac{TP}{TP+FN} \\
\end{gather*}
Thus, we can compute both macro $F_1$-score and weighted $F_1$-score using the following formulas, $N$ denotes the total number of cell types and $n_i$ denotes the number of samples in the $i$-th class:
\begin{gather*}
    macro\:F_1 = \frac{1}{N}\sum_{i=1}^{N}{F_{1}^{(i)}}\\
    weighted\:F_1 = \frac{1}{N}\sum_{i=1}^{N}{n_i*F_{1}^{(i)}} \\
    \textnormal{where} \enspace F_{1}^{(i)}=\frac{2*Precision^{(i)}*Recall^{(i)}}{Precision^{(i)}+Recall^{(i)}}
\end{gather*}

\paragraph{Drug sensitivity prediction.}

The single-cell drug sensitivity prediction task is a binary classification task that aims to predict whether a given cell is sensitive to a particular drug. For this task, we evaluate the effectiveness of the model using the $F_1$-score, which is calculated as follows:
\begin{gather*}
F_{1}=\frac{2*Precision*Recall}{Precision+Recall}
\end{gather*}
where the precision and recall values for the single-cell drug sensitivity prediction task are calculated in the same way as outlined above.

The cell line drug sensitivity prediction task involves predicting the IC50 value in drug sensitivity experiments and is thus a regression task. To evaluate the model’s effectiveness, we use several metrics: Pearson’s correlation coefficient ($\rho_{Y,\hat{Y}}$), $R^2$, root mean squared error (RMSE), and mean absolute error (MAE). Pearson's correlation coefficient reflects the correlation between the predicted and true values, with values ranging from -1 to 1. Values greater than 0 indicate a positive correlation, with values closer to 1 indicating higher correlation. $R^2$ evaluates the goodness of fit of the model, with values between 0 and 1; higher values indicate better fit. RMSE measures the deviation between predicted and true values and is sensitive to outliers in the data; taking the root of RMSE reduces its sensitivity to dimensionality. MAE evaluates the actual magnitude of prediction errors. Smaller values (closer to 0) are better for both RMSE and MAE.

Assuming that the ground truth is $Y=\{y_{1},\, y_{2},\,\cdots\,,y_{n}\}$ and the model's prediction is $\hat{Y}=\{\hat{y}_{1},\, \hat{y}_{2},\,\cdots\,,\hat{y}_{n}\}$, then they are calculated as follows: 
\begin{gather*}
\rho_{Y,\hat{Y}} = \frac{cov(Y,\hat{Y})}{\sigma_{Y}\sigma_{\hat{Y}}} = \frac{E[(Y-\mu_{Y})(\hat{Y}-\mu_{\hat{Y}})]}{\sigma_{Y}\sigma_{\hat{Y}}}, \quad\textnormal{where}\enspace \sigma_{Y}=\sqrt{\frac{1}{n-1}\sum_{i_1}^{n}(Y_i-\bar{Y})^2} \\
R^2 = 1-\frac{\sum_{i}(y_{i}-\hat{y}_{i})^2}{\sum_{i}(y_{i}-\bar{y})^2},\quad \textnormal{where}\enspace \bar{y}=\frac{1}{n}\sum_{i=1}^{n}y_{i}\\
RMSE = \sqrt{\frac{1}{n}\sum_{i=1}^{n}(y_{i}-\hat{y}_{i}))^2} \\
MAE = \frac{1}{n}\sum_{i=1}^{n}|y_{i}-\hat{y}_{i}|\\
\end{gather*}

\subsection{Experiment Configurations for Pertaining and the Downstream Tasks}
\label{sec:appendix6}
The configurations of the model, pre-training, and downstream tasks are shown in Table \ref{Hyperparameters}. Below, we will explain some important details.

\paragraph{Pooling} When using the single-cell gene expression profiles as input to the model, we only select genes with non-zero expression levels. However, when the  model is used for downstream tasks, the model output will be restored to the complete gene sequence. The positions of genes with non-zero expression levels  will be set to the model output. In contrast, the positions of genes with zero expression levels will be set to zero. Subsequently, we can use a simple convolutional neural network and a linear neural network to map the single-cell embeddings to the desired downstream task inputs.

\paragraph{Pre-training} During the MLM  task and the classification task of normal cells and cancer cells, we map the model outputs at corresponding positions(masked genes or \texttt{[CLS]}) to category dimensions using a linear classification head for classification.  In the contrastive learning task, we pool the model's output through a convolutional layer and a linear layer to obtain a 512-dimensional representation.

\paragraph{Downstream Tasks} 
Due to the limitations of the datasets, we only perform single-cell drug sensitivity prediction tasks for a single drug, labeling cells as sensitive or resistant to the drug. Essentially, this task is equivalent to binary classification. For the single-cell drug sensitivity prediction and cell type annotation tasks, we pool the model's output and pass it through several feed-forward neural networks serving as classification heads.
For cell line drug sensitivity prediction, multiple drugs and cell lines are used. We have replicated the drug encoder part of TGDRP \citeapp{TGSAPA}, which encodes drugs in SMILES format into 256-dimensional embeddings. The drug embeddings, along with the 256-dimensional cell embeddings obtained by pooling the model's output, is concatenated and passed through several linear layers to obtain the predicted IC50 values for regression.


\begin{table}[ht]
    \caption{Experiment Configurations}
    \label{Hyperparameters}
    \centering
    \resizebox{0.95\textwidth}{!} {
        \begin{tabular}{lll}
        \toprule
         & \textbf{Hyperparameter} & \textbf{Value} \\
        \midrule
        Model & {\makecell[l]{Number of tokens \\ Feature size \\Number of layers \\ Max sequence length \\ Number of attention heads \\ Dropout}} &  {\makecell[l]{8 \\ 512 \\ 10 \\ 6000 \\ 16 \\ 0.1}} \\
        \midrule
        Pre-training & {\makecell[l]{Size of convolution layer for pooling \\ Size of feed-forward layer for pooling \\ Similarity function \\ Optimizer \\ Scheduler \\ Learning Rate\\Dropout\\Weight decay\\Batch size\\Gradient accumulation}} & {\makecell[l]{(1, 512) \\ 512 \\ Cosine similarity \\ Adam \\ Cosine annealing \\ 0.0001\\ 0.1 \\ 0 \\ 512 \\ 1}} \\
        \midrule
        Cell type annotation & {\makecell[l]{Size of convolution layer for pooling \\ Number of feed-forward layers \\ Feed-forward layer sizes \\ Activation function \\ Optimizer \\ Learning Rate\\Dropout\\Weight decay\\Batch size\\Gradient accumulation}} & {\makecell[l]{(1, 512) \\ 3 \\ 512, 128, 11/13 \\ ReLU \\ Adam \\ 0.0001\\ 0.1 \\ 0.0001 \\ 32 \\ 8}} \\
        \midrule
        {\makecell[l]{Single-cell drug\\sensitivity prediction}} & {\makecell[l]{Size of convolution layer for pooling \\ Number of feed-forward layers\\Feed-forward layer sizes\\Activation function\\Optimizer\\Learning Rate\\Dropout\\Weight decay\\Batch size\\Gradient accumulation}} & {\makecell[l]{(1, 512)\\3 \\ 512, 32, 2\\ LeakyReLU \\Adam \\ 0.0001 \\0.1 \\ 0.001 \\ 4 \\ 4 }} \\
        \midrule
        {\makecell[l]{Cell line drug\\sensitivity prediction}} & {\makecell[l]{Size of convolution layer for pooling \\ Number of feed-forward layers for cell \\ Sizes of feed-forward layers for cell  \\ Activation function for cell \\ Cell feature size \\ Drug feature size \\ Number of feed-forward layers for regression \\ Sizes of feed-forward layers for regression  \\ Activation function for regression \\ Optimizer \\ Learning Rate\\Dropout\\Weight decay\\Batch size\\Gradient accumulation}} & {\makecell[l]{(1, 512) \\ 2 \\ 1024, 256 \\ ReLU \\ 256 \\ 256 \\ 3 \\ 512, 512, 1 \\ ELU \\ Adam \\ 0.0001\\ 0.2 \\ 0 \\ 16 \\ 1}} \\
        \bottomrule
        \end{tabular}
    }
\end{table}

\clearpage
\small
\bibliographystyleapp{unsrt}
\bibliographyapp{ref_appendix}


\end{document}